\documentclass[reprint,showpacs,preprintnumbers,nofootinbib,amsmath,amssymb,aps,prd,floatfix]{revtex4-1}
\usepackage{hyperref}%
\usepackage[english]{babel}
\usepackage{graphicx}
\date{\today}

\usepackage{color}
\definecolor{mygreen}{rgb}{1, 0, 0}

\begin{document}

\title{Big Rip: heating by Hawking radiation and a possible\\
connection to conformal cyclic cosmology}
\author{Rafael Ruggiero}\email{ruggiero3n1@gmail.com}

\affiliation{S\~ao Paulo, Brazil}



\begin{abstract}
In the Big Rip cosmological scenario, a FRW universe containing dark energy with $w < -1$ in its equation of state (phantom energy) expands in such way that, in a finite time span, the scale factor diverges to infinity and the size of the cosmic horizon goes to zero. Here we revisit this scenario in light of the fact that Hawking radiation is expected to be generated at the apparent horizon of a FRW universe, and show that the energy density and temperature of that radiation both diverge at the Big Rip. We then use this fact to propose a new variant of Penrose's conformal cyclic cosmology model in which the future spacetime metric becomes conformally invariant at the Big Rip instead of in the remote future of a de Sitter universe; this removes the need for mass decay from the model and makes it consistent with current physical laws.
\end{abstract}


\maketitle

\section{Introduction}\label{sec:I}

The ultimate fate of the Universe is a topic of great philosophical significance.
In order to speculate about what will take place, it is crucial to understand what
factors are involved in its expansion. In particular, the future evolution strongly
depends on the exact nature of dark energy.

In a generalized formulation, dark energy is a fluid with equation of state $P = w \rho$.
Current observations suggest that $w$ is close to $-1$ (e.g. \cite{planck}), and an often adopted
base scenario is that its value is exactly $-1$, 
making dark energy a negative pressure fluid 
that can be more simply interpreted as a cosmological constant in the Einstein equations, since the behavior
is exactly the same. Dark energy with $w < -1$ in its equation of state receives the name of \emph{phantom energy} \cite{phantomoriginal},
and it results in an extreme future expansion that diverges in a finite time span and causes the size of the cosmic horizon to go to zero.
That would
disrupt all structures, from galaxies to molecules and atoms, hence why this scenario is called \emph{Big Rip} \cite{bigrip}.

The metric for an homogeneous and isotropic universe
is the Friedmann-Robertson-Walker (FRW) metric.
One important property of a FRW universe is that Hawking radiation
is expected to be generated at its apparent horizon. This has been demonstrated in
\cite{frwuniverse} using the so called \emph{tunneling approach} \cite{tunneling}, in which 
the process of particle creation at the horizon is conceptualized as the
tunneling of particles from beyond it.
This result generalizes the previous finding by Gibbons and Hawking that 
the apparent horizon of a de Sitter universe is expected to radiate \cite{cosmohawking},
obtained just a few years after Hawking first considered the problem of particle creation
at the event horizon of black holes \cite{hawking}. 

The temperature of the Hawking radiation generated at the apparent horizon of a FRW universe
is inversely proportional to the radius $\widetilde{r}_A$ of the horizon, and is given by
\cite{Cai:2005ra,frwuniverse}:
\begin{equation} \label{eq:T}
T = \frac{\hbar c}{k_B} \frac{1}{2 \pi \widetilde{r}_A}.
\end{equation}
This result holds for any spatial curvature. In the case of a flat universe ($k = 0$), the radius of the horizon is simply \cite{Cai:2005ra}
\begin{equation} \label{eq:radius}
\widetilde{r}_A = \frac{c}{H},
\end{equation}
where $H \equiv \dot{a}/{a}$ is the Hubble parameter.

As intriguing as the fate of the universe is its origin. The question remains as
to whether the Big Bang was an ultimate beginning or if it was preceded by something.
Cyclic cosmological models hold to the latter view; a prominent such model has been proposed
by Roger Penrose and is called \emph{Conformal Cyclic Cosmology} (CCC, \cite{Penrose2011,Penrose2012}). The basic
idea of this model is that, as long all mass in the universe asymptotically
decays in an unspecified way in the remote future, the spacetime metric will
eventually become conformally invariant, just like it was at the Big Bang
when radiation was the only relevant energy component, making it possible to
reinterpret the final state of the universe as something equivalent to its
initial state. In this framework,
each iteration of the cyclic universe is referred to as an \emph{aeon},
and infinitely many of these aeons are assumed to take place in sequence, without
a beginning or an end.
Tests of this model based on attempting to find gravitational signatures
of a previous aeon
on the CMB have been carried out with positive
results \cite{cmb1,cmb2}, but their statistical relevance has been disputed \cite{Jow_2020}.

In this paper, we revisit the Big Rip scenario considering the Hawking radiation that is generated
at the apparent horizon of a FRW universe containing phantom energy. 
In Sec.~\ref{sec:II},
we show that the temperature and energy
density of that radiation both diverge at the Big Rip. In
Sec.~\ref{sec:III}, we propose a variant of the CCC model in which the final, conformally invariant
limit of the universe is taken to be the Big Rip instead of the remote future of a de Sitter universe.
We then discuss and contextualize our results in Sec.~\ref{sec:IV},
and summarize in Sec.~\ref{sec:V}.

\section{Heating by Hawking radiation}\label{sec:II}

To calculate the heating of a phantom universe by Hawking radiation, we start from the
Friedmann equation. For an universe similar to ours at the present moment, 
assumed to be flat and with matter and dark energy being the only relevant energy components,
it can be written as \cite{bigrip}:
\begin{equation} \label{eq:fried1}
\frac{1}{H_0^2} \frac{\dot{a}^2}{a^2} = \Omega_{m,0}\,a^{-3} + (1-\Omega_{\mathrm{m},0})\,a^{-3(1+w)},
\end{equation}
where $a$ is the scale factor, $H_0$ is the current Hubble parameter and $\Omega_{m,0}$ is
the current matter density parameter. 

\begin{figure}[h]
    \includegraphics[width=\columnwidth]{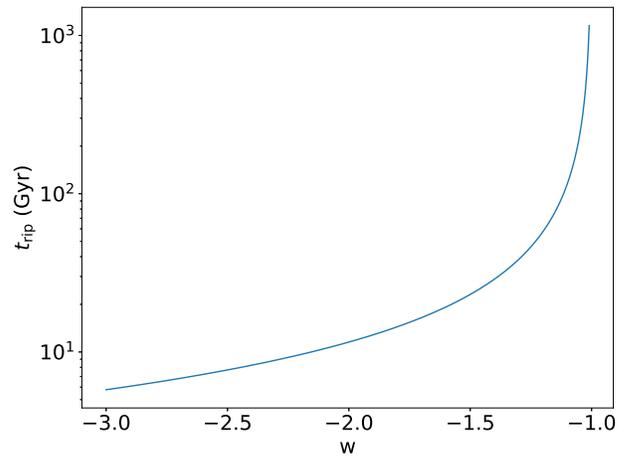}
    \caption{Time to the Big Rip as a function of the dark energy equation of state parameter $w$, calculated
    using $H_0 =  0.069$ Gyr$^{-1}$ and $\Omega_{\mathrm{m},0} = 0.3$.}
    \label{fig:trip}
\end{figure}

If $w < -1$, then the term $\Omega_{m,0} \,a^{-3}$ will eventually become much smaller
than the second one, so that the equation will become simply
\begin{equation} \label{eq:fried2}
\frac{1}{H_0^2} \frac{\dot{a}^2}{a^2} =  (1-\Omega_{\mathrm{m},0})\,a^{-3(1+w)}.
\end{equation}
The solution for this equation is of the form $a(t) \propto (A - B t)^{-C}$, where
$A$ and $C$ are positive numbers determined by $w$, and $B$ is a positive number
determined by $H_0$ and $\Omega_{\mathrm{m},0}$. This solution diverges when $A - B t = 0$;
an explicit calculation shows that this happens for 
\begin{equation}
t_{\mathrm{rip}} = - \frac{2}{3 (1 + w) H_0 \sqrt{1 - \Omega_{\mathrm{m},0}}},
\end{equation}
and that $H$ also diverges at this moment. Figure \ref{fig:trip} illustrates 
some values of $t_{\mathrm{rip}}$ as a function of $w$.

The Hawking radiation temperature given by Eq.~\ref{eq:T} has been shown in \cite{frwuniverse}
to apply to any universe
obeying the FRW metric, a special case of which is the universe containing phantom
energy that we are considering. Since this temperature is proportional to $H$
(as a consequence of Eq.~\ref{eq:radius}), we conclude that it will diverge at the
Big Rip. The same is true 
for the energy density
associated to that temperature, which assuming thermal equilibrium is given by the Stefan-Boltzmann law for black-body radiation:
\begin{eqnarray} \label{eq:energy}
U &=& \frac{4 \sigma}{c} T^4  \\
  &=& \frac{4 \sigma}{c} \left( \frac{\hbar c}{k_B} \frac{1}{2 \pi \widetilde{r}_A} \right)^4.
\end{eqnarray}

We started the discussion assuming that dark energy would be
the only relevant energy component in the universe, and ended up concluding that a new energy component (Hawking
radiation) appears as a result. The back-reaction that results from this has
been shown in \cite{hawkinginflation1,hawkinginflation2} to be a viable
mechanism of inflation, which will be relevant for us 
in Sec.~\ref{sec:III}. For the remainder of this paper, we are going to
assume that the cosmological back-reaction of Hawking radiation
will only accelerate the expansion of the
phantom universe, without any qualitative change to its outcome.

\section{Conformal cyclic cosmology}\label{sec:III}

Now that we have established that the energy density of Hawking radiation diverges at the Big Rip,
we turn to considering how this might establish a connection between the Big Rip scenario and
conformal cyclic cosmology.
The main observation to be made is that the energy density of matter relative to radiation
goes to zero at the Big Rip, causing the spacetime metric to become conformally invariant at that
moment. With this, the procedure of ``squashing down'' a divergent metric that is employed
in the CCC model can also be applied in this case. It consists of reinterpreting the
metric using a smooth positive scalar field $\Omega$:
\begin{equation}
g_\mathrm{\mu \nu} \rightarrow \Omega^2 g_\mathrm{\mu \nu}.
\end{equation}
The conformal factor $\Omega$ is taken to be infinitesimal, and the result of the transformation
is to bring a future divergent scale factor to a finite
size. This may sound like an unreasonable procedure at first glance, but the removal of isotropic spacetime singularities through
conformal rescalings is mathematically well defined, as outlined in \cite{Tod_2003}.
A side effect of this transformation is to make the energy density of phantom energy insignificant,
given that it is proportional to $a^{-3(1+w)}$ while $a$ is being made small,
thus bringing the universe to a state completely dominated by radiation that would correspond to
a new Big Bang, exactly as in CCC. 

This way, it can be seen that the Big Rip scenario leads to a variant of the CCC model, but with some differences. The first
is that the need for all particles in the universe to lose their masses in the future is removed, since the
spacetime metric becomes conformally invariant due to the radiation energy density increasing instead of
the matter energy density decreasing. The second is that the age of the universe at the end of each aeon is finite in
this variation, whereas it is infinite in the original formulation.

Regarding inflation, in the CCC model it is loosely assumed that the exponential expansion of a previous
aeon will play the role of inflation without the need for any ad hoc inflaton field. In our variant of the model, the
expansion prior to the Big Rip could perhaps be taken to play the same role,
but we consider it more interesting to adhere to the view of inflation proposed in \cite{hawkinginflation1,hawkinginflation2},
in which it is driven by the Hawking radiation generated at the apparent horizon.
The basic idea is that a very dense universe with a fast
expansion rate and a small apparent horizon sees a high energy density of Hawking radiation,
which causes the expansion to self-reinforce and become exponential, but in a way that
does not go on forever:
once the original density of the universe gets diluted
enough by the expansion, Hawking radiation becomes insignificant and
the exponential expansion ends.  This has been quantitatively shown to
be a viable mechanism for inflation, and it is one that is compatible
with our framework, given that a high energy density in the beginning
of a new aeon is an outcome of it.

\section{Discussion}\label{sec:IV}

Relative to the original CCC formulation, our version has the advantage of
not requiring all the massive fermions and massive charged particles in the universe
to disappear in the future. In fact, not even
the evaporation of the black holes in the universe is a requirement. But it
comes with the need for a new ingredient, namely the presence of phantom energy.
 Current observations are in agreement with this requirement, 
and will remain this way
as long as $w = -1$ remains a valid option, since error bars
will always exist and any value $w < -1$, no matter how close to $-1$, 
suffices to cause a future Big Rip, as pointed out in \cite{bigrip}.
All that is changed is how far into the future this moment will be. Perhaps
the biggest disadvantage of phantom energy relative to the $w = -1$ base
scenario is that it cannot be interpreted
as a cosmological constant. So in order for its 
effects to be seen without the introduction of an ad hoc energy component
in the universe, a more sophisticated modification of General Relativity
would have to be developed. It is worth noting that an equation of state
parameter $w$ that remains constant in time is a
simplifying assumption; the Big Rip is also bound to happen e.g. for a time 
varying dark energy model which satisfies $w < -1$ at all times.

Regarding observational tests of our model, we consider
that the same ones that apply to the original CCC model also apply to it, namely 
looking for signatures of a previous aeon on the CMB \cite{cmb1,cmb2} and possibly on the gravitational
wave background of the universe, which has been shown to be sensitive to the
primordial density fluctuations \cite{PhysRevD.75.123518}. Both of these
lie on the fact that gravitational radiation is able to cross the divergent
spacetime boundary in the CCC framework \cite{Penrose2012}.
The predictions could turn out to have discerning characteristics in our case
because, since the previous aeon would have ended in a finite time span and in a more
abrupt way, the irregularities in its final energy distribution would have been presumably different
than in the more gradual scale factor divergence of standard CCC.

\section{Conclusion}\label{sec:V}

In this paper, we have shown that the Hawking radiation generated at the apparent horizon
of a FRW universe containing phantom energy (dark energy with $w < -1$ in its
equation of state) diverges both in temperature and energy density at the Big Rip.
Then we used this fact to propose a variant of the conformal cyclic cosmology model based on the observation that
the energy density of matter relative to radiation will become insignificant at the Big Rip, making the spacetime metric
conformally invariant at that moment. This variant makes the model more compatible with known physical laws, since the need for mass decay is removed from it. The dark energy requirements are also consistent with the best cosmological parameter estimations available today, given that values of $w$ smaller than $-1$ are within their confidence intervals.
Tying this framework to a model of inflation based on Hawking radiation, we believe that a fully self-consistent cyclic cosmological model can be obtained.

\bibliography{cosmo}

\begin{thebibliography}{17}%
\makeatletter
\providecommand \@ifxundefined [1]{%
 \@ifx{#1\undefined}
}%
\providecommand \@ifnum [1]{%
 \ifnum #1\expandafter \@firstoftwo
 \else \expandafter \@secondoftwo
 \fi
}%
\providecommand \@ifx [1]{%
 \ifx #1\expandafter \@firstoftwo
 \else \expandafter \@secondoftwo
 \fi
}%
\providecommand \natexlab [1]{#1}%
\providecommand \enquote  [1]{``#1''}%
\providecommand \bibnamefont  [1]{#1}%
\providecommand \bibfnamefont [1]{#1}%
\providecommand \citenamefont [1]{#1}%
\providecommand \href@noop [0]{\@secondoftwo}%
\providecommand \href [0]{\begingroup \@sanitize@url \@href}%
\providecommand \@href[1]{\@@startlink{#1}\@@href}%
\providecommand \@@href[1]{\endgroup#1\@@endlink}%
\providecommand \@sanitize@url [0]{\catcode `\\12\catcode `\$12\catcode
  `\&12\catcode `\#12\catcode `\^12\catcode `\_12\catcode `\%12\relax}%
\providecommand \@@startlink[1]{}%
\providecommand \@@endlink[0]{}%
\providecommand \url  [0]{\begingroup\@sanitize@url \@url }%
\providecommand \@url [1]{\endgroup\@href {#1}{\urlprefix }}%
\providecommand \urlprefix  [0]{URL }%
\providecommand \Eprint [0]{\href }%
\providecommand \doibase [0]{http://dx.doi.org/}%
\providecommand \selectlanguage [0]{\@gobble}%
\providecommand \bibinfo  [0]{\@secondoftwo}%
\providecommand \bibfield  [0]{\@secondoftwo}%
\providecommand \translation [1]{[#1]}%
\providecommand \BibitemOpen [0]{}%
\providecommand \bibitemStop [0]{}%
\providecommand \bibitemNoStop [0]{.\EOS\space}%
\providecommand \EOS [0]{\spacefactor3000\relax}%
\providecommand \BibitemShut  [1]{\csname bibitem#1\endcsname}%
\let\auto@bib@innerbib\@empty
\bibitem [{\citenamefont {Collaboration}(2018)}]{planck}%
  \BibitemOpen
  \bibfield  {author} {\bibinfo {author} {\bibfnamefont {P.}~\bibnamefont
  {Collaboration}},\ }\href@noop {} {\enquote {\bibinfo {title} {Planck 2018
  results. vi. cosmological parameters},}\ } (\bibinfo {year} {2018}),\ \Eprint
  {http://arxiv.org/abs/1807.06209} {arXiv:1807.06209 [astro-ph.CO]}
  \BibitemShut {NoStop}%
\bibitem [{\citenamefont {Caldwell}(2002)}]{phantomoriginal}%
  \BibitemOpen
  \bibfield  {author} {\bibinfo {author} {\bibfnamefont {R.}~\bibnamefont
  {Caldwell}},\ }\href {\doibase 10.1016/s0370-2693(02)02589-3} {\bibfield
  {journal} {\bibinfo  {journal} {Physics Letters B}\ }\textbf {\bibinfo
  {volume} {545}},\ \bibinfo {pages} {23–29} (\bibinfo {year}
  {2002})}\BibitemShut {NoStop}%
\bibitem [{\citenamefont {Caldwell}\ \emph {et~al.}(2003)\citenamefont
  {Caldwell}, \citenamefont {Kamionkowski},\ and\ \citenamefont
  {Weinberg}}]{bigrip}%
  \BibitemOpen
  \bibfield  {author} {\bibinfo {author} {\bibfnamefont {R.~R.}\ \bibnamefont
  {Caldwell}}, \bibinfo {author} {\bibfnamefont {M.}~\bibnamefont
  {Kamionkowski}}, \ and\ \bibinfo {author} {\bibfnamefont {N.~N.}\
  \bibnamefont {Weinberg}},\ }\href {\doibase 10.1103/PhysRevLett.91.071301}
  {\bibfield  {journal} {\bibinfo  {journal} {Phys. Rev. Lett.}\ }\textbf
  {\bibinfo {volume} {91}},\ \bibinfo {pages} {071301} (\bibinfo {year}
  {2003})}\BibitemShut {NoStop}%
\bibitem [{\citenamefont {{Cai}}\ \emph {et~al.}(2009)\citenamefont {{Cai}},
  \citenamefont {{Cao}},\ and\ \citenamefont {{Hu}}}]{frwuniverse}%
  \BibitemOpen
  \bibfield  {author} {\bibinfo {author} {\bibfnamefont {R.-G.}\ \bibnamefont
  {{Cai}}}, \bibinfo {author} {\bibfnamefont {L.-M.}\ \bibnamefont {{Cao}}}, \
  and\ \bibinfo {author} {\bibfnamefont {Y.-P.}\ \bibnamefont {{Hu}}},\ }\href
  {\doibase 10.1088/0264-9381/26/15/155018} {\bibfield  {journal} {\bibinfo
  {journal} {Classical and Quantum Gravity}\ }\textbf {\bibinfo {volume}
  {26}},\ \bibinfo {eid} {155018} (\bibinfo {year} {2009})},\ \Eprint
  {http://arxiv.org/abs/0809.1554} {arXiv:0809.1554 [hep-th]} \BibitemShut
  {NoStop}%
\bibitem [{\citenamefont {Parikh}\ and\ \citenamefont
  {Wilczek}(2000)}]{tunneling}%
  \BibitemOpen
  \bibfield  {author} {\bibinfo {author} {\bibfnamefont {M.~K.}\ \bibnamefont
  {Parikh}}\ and\ \bibinfo {author} {\bibfnamefont {F.}~\bibnamefont
  {Wilczek}},\ }\href {\doibase 10.1103/PhysRevLett.85.5042} {\bibfield
  {journal} {\bibinfo  {journal} {Phys. Rev. Lett.}\ }\textbf {\bibinfo
  {volume} {85}},\ \bibinfo {pages} {5042} (\bibinfo {year}
  {2000})}\BibitemShut {NoStop}%
\bibitem [{\citenamefont {Gibbons}\ and\ \citenamefont
  {Hawking}(1977)}]{cosmohawking}%
  \BibitemOpen
  \bibfield  {author} {\bibinfo {author} {\bibfnamefont {G.~W.}\ \bibnamefont
  {Gibbons}}\ and\ \bibinfo {author} {\bibfnamefont {S.~W.}\ \bibnamefont
  {Hawking}},\ }\href {\doibase 10.1103/PhysRevD.15.2738} {\bibfield  {journal}
  {\bibinfo  {journal} {Phys. Rev. D}\ }\textbf {\bibinfo {volume} {15}},\
  \bibinfo {pages} {2738} (\bibinfo {year} {1977})}\BibitemShut {NoStop}%
\bibitem [{\citenamefont {Hawking}(1975)}]{hawking}%
  \BibitemOpen
  \bibfield  {author} {\bibinfo {author} {\bibfnamefont {S.~W.}\ \bibnamefont
  {Hawking}},\ }\href {https://projecteuclid.org:443/euclid.cmp/1103899181}
  {\bibfield  {journal} {\bibinfo  {journal} {Comm. Math. Phys.}\ }\textbf
  {\bibinfo {volume} {43}},\ \bibinfo {pages} {199} (\bibinfo {year}
  {1975})}\BibitemShut {NoStop}%
\bibitem [{\citenamefont {Cai}\ and\ \citenamefont {Kim}(2005)}]{Cai:2005ra}%
  \BibitemOpen
  \bibfield  {author} {\bibinfo {author} {\bibfnamefont {R.-G.}\ \bibnamefont
  {Cai}}\ and\ \bibinfo {author} {\bibfnamefont {S.~P.}\ \bibnamefont {Kim}},\
  }\href {\doibase 10.1088/1126-6708/2005/02/050} {\bibfield  {journal}
  {\bibinfo  {journal} {JHEP}\ }\textbf {\bibinfo {volume} {02}},\ \bibinfo
  {pages} {050} (\bibinfo {year} {2005})},\ \Eprint
  {http://arxiv.org/abs/hep-th/0501055} {arXiv:hep-th/0501055} \BibitemShut
  {NoStop}%
\bibitem [{\citenamefont {{Penrose}}(2011)}]{Penrose2011}%
  \BibitemOpen
  \bibfield  {author} {\bibinfo {author} {\bibfnamefont {R.}~\bibnamefont
  {{Penrose}}},\ }\href@noop {} {\emph {\bibinfo {title} {Cycles of Time: An
  Extraordinary New View of the Universe}}}\ (\bibinfo  {publisher} {Vintage},\
  \bibinfo {address} {London},\ \bibinfo {year} {2011})\BibitemShut {NoStop}%
\bibitem [{\citenamefont {{Penrose}}(2012)}]{Penrose2012}%
  \BibitemOpen
  \bibfield  {author} {\bibinfo {author} {\bibfnamefont {R.}~\bibnamefont
  {{Penrose}}},\ }in\ \href {\doibase 10.1063/1.4727997} {\emph {\bibinfo
  {booktitle} {Ameri. Inst. of Phy. Conf. Series}}},\ \bibinfo {series}
  {American Institute of Physics Conference Series}, Vol.\ \bibinfo {volume}
  {1446},\ \bibinfo {editor} {edited by\ \bibinfo {editor} {\bibfnamefont
  {J.}~\bibnamefont {{Kouneiher}}}, \bibinfo {editor} {\bibfnamefont
  {C.}~\bibnamefont {{Barbachoux}}}, \bibinfo {editor} {\bibfnamefont
  {T.}~\bibnamefont {{Masson}}}, \ and\ \bibinfo {editor} {\bibfnamefont
  {D.}~\bibnamefont {{Vey}}}}\ (\bibinfo {year} {2012})\ pp.\ \bibinfo {pages}
  {233--243}\BibitemShut {NoStop}%
\bibitem [{\citenamefont {Gurzadyan}\ and\ \citenamefont
  {Penrose}(2013)}]{cmb1}%
  \BibitemOpen
  \bibfield  {author} {\bibinfo {author} {\bibfnamefont {V.}~\bibnamefont
  {Gurzadyan}}\ and\ \bibinfo {author} {\bibfnamefont {R.}~\bibnamefont
  {Penrose}},\ }\href {\doibase 10.1140/epjp/i2013-13022-4} {\bibfield
  {journal} {\bibinfo  {journal} {The E. P. J. Plus}\ }\textbf {\bibinfo
  {volume} {128}} (\bibinfo {year} {2013}),\
  10.1140/epjp/i2013-13022-4}\BibitemShut {NoStop}%
\bibitem [{\citenamefont {An}\ \emph {et~al.}(2018)\citenamefont {An},
  \citenamefont {Meissner}, \citenamefont {Nurowski},\ and\ \citenamefont
  {Penrose}}]{cmb2}%
  \BibitemOpen
  \bibfield  {author} {\bibinfo {author} {\bibfnamefont {D.}~\bibnamefont
  {An}}, \bibinfo {author} {\bibfnamefont {K.~A.}\ \bibnamefont {Meissner}},
  \bibinfo {author} {\bibfnamefont {P.}~\bibnamefont {Nurowski}}, \ and\
  \bibinfo {author} {\bibfnamefont {R.}~\bibnamefont {Penrose}},\ }\href@noop
  {} {\enquote {\bibinfo {title} {Apparent evidence for hawking points in the
  cmb sky},}\ } (\bibinfo {year} {2018}),\ \Eprint
  {http://arxiv.org/abs/1808.01740} {arXiv:1808.01740 [astro-ph.CO]}
  \BibitemShut {NoStop}%
\bibitem [{\citenamefont {Jow}\ and\ \citenamefont {Scott}(2020)}]{Jow_2020}%
  \BibitemOpen
  \bibfield  {author} {\bibinfo {author} {\bibfnamefont {D.~L.}\ \bibnamefont
  {Jow}}\ and\ \bibinfo {author} {\bibfnamefont {D.}~\bibnamefont {Scott}},\
  }\href {\doibase 10.1088/1475-7516/2020/03/021} {\bibfield  {journal}
  {\bibinfo  {journal} {J. of Cosmol. and Astrop. Physics}\ }\textbf {\bibinfo
  {volume} {2020}},\ \bibinfo {pages} {021–021} (\bibinfo {year}
  {2020})}\BibitemShut {NoStop}%
\bibitem [{\citenamefont {Modak}\ and\ \citenamefont
  {Singleton}(2012)}]{hawkinginflation1}%
  \BibitemOpen
  \bibfield  {author} {\bibinfo {author} {\bibfnamefont {S.~K.}\ \bibnamefont
  {Modak}}\ and\ \bibinfo {author} {\bibfnamefont {D.}~\bibnamefont
  {Singleton}},\ }\href {\doibase 10.1103/PhysRevD.86.123515} {\bibfield
  {journal} {\bibinfo  {journal} {Phys. Rev. D}\ }\textbf {\bibinfo {volume}
  {86}},\ \bibinfo {pages} {123515} (\bibinfo {year} {2012})}\BibitemShut
  {NoStop}%
\bibitem [{\citenamefont {Modak}\ and\ \citenamefont
  {Singleton}(2013)}]{hawkinginflation2}%
  \BibitemOpen
  \bibfield  {author} {\bibinfo {author} {\bibfnamefont {S.~K.}\ \bibnamefont
  {Modak}}\ and\ \bibinfo {author} {\bibfnamefont {D.}~\bibnamefont
  {Singleton}},\ }\href {\doibase 10.1063/1.4791745} {\bibfield  {journal}
  {\bibinfo  {journal} {AIP Conference Proceedings}\ }\textbf {\bibinfo
  {volume} {1514}},\ \bibinfo {pages} {150} (\bibinfo {year} {2013})},\ \Eprint
  {http://arxiv.org/abs/https://aip.scitation.org/doi/pdf/10.1063/1.4791745}
  {https://aip.scitation.org/doi/pdf/10.1063/1.4791745} \BibitemShut {NoStop}%
\bibitem [{\citenamefont {Tod}(2003)}]{Tod_2003}%
  \BibitemOpen
  \bibfield  {author} {\bibinfo {author} {\bibfnamefont {K.~P.}\ \bibnamefont
  {Tod}},\ }\href {\doibase 10.1088/0264-9381/20/3/309} {\bibfield  {journal}
  {\bibinfo  {journal} {Classical and Quantum Gravity}\ }\textbf {\bibinfo
  {volume} {20}},\ \bibinfo {pages} {521} (\bibinfo {year} {2003})}\BibitemShut
  {NoStop}%
\bibitem [{\citenamefont {Ananda}\ \emph {et~al.}(2007)\citenamefont {Ananda},
  \citenamefont {Clarkson},\ and\ \citenamefont {Wands}}]{PhysRevD.75.123518}%
  \BibitemOpen
  \bibfield  {author} {\bibinfo {author} {\bibfnamefont {K.~N.}\ \bibnamefont
  {Ananda}}, \bibinfo {author} {\bibfnamefont {C.}~\bibnamefont {Clarkson}}, \
  and\ \bibinfo {author} {\bibfnamefont {D.}~\bibnamefont {Wands}},\ }\href
  {\doibase 10.1103/PhysRevD.75.123518} {\bibfield  {journal} {\bibinfo
  {journal} {Phys. Rev. D}\ }\textbf {\bibinfo {volume} {75}},\ \bibinfo
  {pages} {123518} (\bibinfo {year} {2007})}\BibitemShut {NoStop}%
\end{thebibliography}%

\end{document}